\def   \ni {\noindent}

\def   \Msun {M$_{\odot}$\ }

\def   \hMpc {{{$h^{-1}$}Mpc}\ }
\def   \Mpc {{Mpc}}
\def   \ea {{et\thinspace al.}\ }
 
\def   \ssk {\vskip  5truept}

\def   \bsk {\vskip 15truept}

\def   \newline {\hfil\break}

\magnification=1000
\hsize 5truein
\vsize 8.5truein
\topskip 1cm
\font\abstract=cmr8

\font\text=cmr10
\font\affiliation=cmssi10
\font\author=cmss10

\font\title=cmssbx10 scaled\magstep2

\def\ref{\par\noindent\hangindent 15pt}

\null
\vskip 2.5truecm
\baselineskip = 12pt

{\title\ni Redshift dependence of clustering in numerical simu\-lations}
\bsk \bsk
{\author\ni VOLKER M\"ULLER}
\bsk
{\affiliation\ni Astrophysical Institute Potsdam}
\bsk
\bsk
\baselineskip = 9pt
{\abstract 
\ni
High resolution $N$-body simulations for variants of the CDM model are used to
derive the redshift dependence of galaxy formation and of galaxy correlation
functions.  The reconstructed power spectra and clustering properties provide a
sensible test for the underlying power spectra.  We compare our simulations
with new results of the analysis of recent redshift surveys. The results are
used to discuss the model of primordially broken scale invariant perturbations
(BSI).}
\bsk
\baselineskip = 12pt

{\text                                                    
\ni 1. INTRODUCTION

{\sl Biased} structure formation in a flat universe dominated by cold dark
matter (CDM) has difficulties in describing the formation of large-scale
structure in the universe, it delivers too small large-scale power for getting
the microwave background anisotropies, large scale velocity fields, and
cluster-cluster correlations.  On the other hand, {\sl unbiased} (COBE
normalised) CDM seems to have too much small scale power for describing the
galaxy clustering and small scale velocity fields.  This was our motivation for
discussing double inflationary models which provide perturbation spectra with
enhanced power at large scales (Broken Scale Invariant perturbations -- BSI,
Gottl\"ober et al.  1991).  The redshift dependence of gravitational clustering
of the new theory in comparison with COBE-normalised standard CDM enlightens its
pros and cons.

To this aim we performed a series of high-resolution PM simulations, with a
model for the thermodynamic evolution of baryons (cp.  part III and Kates et al.
1995 for details and chosen parameters).  The superposition of simulation boxes
with different scales allows the study of structure formation over a wide range
of cosmic scales.  The 'thermodynamic' properties are used as an additional
indicator for dissipative processes in the high density regions, and therefore,
for the time dependence of the condensation of cosmic structures.  It is
interesting that the filamentary structure becomes visible in the distribution
of both 'cold' particles and warm ($\approx 10^6$~K) gas.

We use density peaks in the {\sl cold} matter for identifying 'galaxy' halos
with a reasonable mass spectrum, and we compare their clustering properties and
power spectra with data from large galaxy catalogues.  Next we will discuss the
redshift evolution of the two-point correlation function using peak-selected
mass particles.  Finally, we compare our results with data from observations,
and we draw some conclusion.

\smallskip
\ni II. DOUBLE INFLATION SCENARIO

Double inflation is an old idea which requires for example two basic scalar
fields in the fundamental theory (Starobinsky 1985).  We derived in detail the
conditions for the occurrence of two inflationary stages separated by an
intermediate dust-like phase (Gottl\"ober et al.  1991).  This is necessary for
getting a sharp break in the primordial power spectrum.  Further, we select the
parameters so that this break lies at the comoving scale corresponding
approximately to the turn around radius of galaxy clusters.  Then we have a
typical scale which separates 'large' and 'small' cosmic structures.  A
description of the power of primordial potential perturbations $P_{\Phi}(k)$ is
given by the approximation (it does not describe the typical oscillations
generated during the intermediate power law stage),
$$ k^{3} P_{\Phi}(k) = \cases{4.2\times 10^{-6} [\log{(k_s/k)}]^{0.6} +
                                   4.7\times 10^{-6} & for $k < k_s$ \cr 
                              9.4\times 10^{-8} \log{(k_f/k)} & for $k
                                   > k_s$ \cr } 
$$ 
Here, $k_s = (2 \pi /24)$\hMpc, $k_f = e^{56}$\hMpc.  We considered a 
cosmological model with the dimensionless density parameter $\Omega =1$,
dominated by cold dark matter (CDM), and we use a dimensionless Hubble constant
$h=0.5$.  The initial power spectrum rescaled by linear theory to redshift $z=0$
is shown by the dash-dotted line in Fig.  1.  For comparison we show the CDM
curve as dotted line.

\midinsert
\vskip 1.5truecm
\par\noindent
\vglue7.3cm
\includegraphics{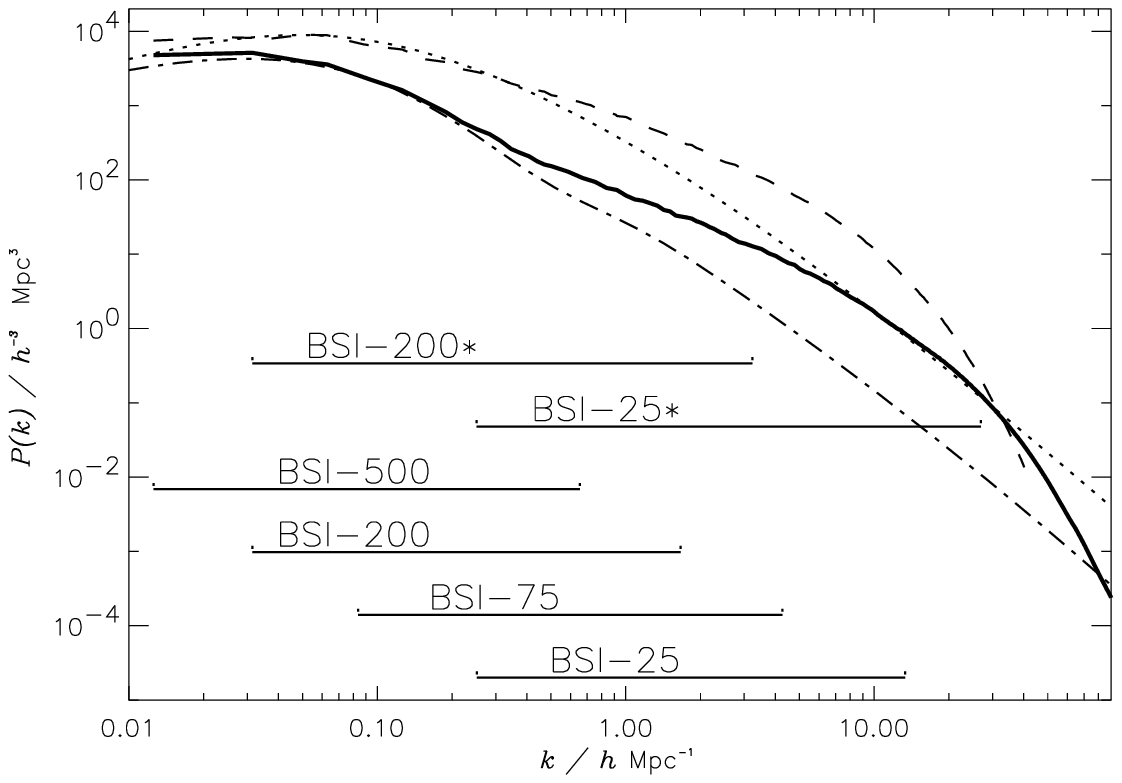}
\leftskip=1truecm
\rightskip=1truecm
\noindent
{\abstract FIGURE 1. Linear and non-linear BSI (dash-dotted and solid lines) and 
CDM (dotted and dashed lines) power spectra of density fluctuations at $z=0$.}
\vskip -0.3truecm
\leftskip=0truecm
\rightskip=0truecm
\endinsert

An alternative model which uses a feature in the scalar field potential was
discussed by Semig and M\"uller (1995).  It describes the transition of a
self-interacting to a massive scalar field, in consequence it leads to a
valley-like shape of the primordial potential spectrum.  All the power spectra
used in the simulations are normalised at large scales by the measured
anisotropy of the microwave background radiation (cp.  Gottl\"ober and M\"ucket
1993).

\smallskip
\ni III. N-BODY SIMULATIONS AND GALAXY IDENTIFICATION

At presence no theoretical description or simulation scheme exhaust all the
various effects of structure formation in the universe.  For testing our
envisaged model we have to study both the large scale structure formation, i.e.
the formation of cluster and superclusters of galaxies, and the formation of the
different types of galaxies as the basic building blocks of all large cosmic
structures.

To this end, we perform a set of simulations with box sizes ranging from 25
\hMpc for having enough resolution at galactic scales, to 75 \hMpc and 200 \hMpc
for studying galaxy clustering, and to 500 \hMpc for the super-large scale
structure (in Fig.  1 the covered wavenumber range is schematically indicated).
We use the PM code of Kates, Kotok, $\&$ Klypin (1991), taking for BSI in the
highest resolution $512^3$ cells and $256^3$ particles (simulation names with
asterisk in Fig.  1) and a set of simulations with half the resolution for
comparing BSI and CDM (Kates \ea 1995).  The solid line in Fig.  1 shows the
combined non-linear spectrum of the dark matter after evolving from redshift
$z=25$ to the presence.  The higher non-linear power of the corresponding
CDM-simulations is shown by the dashed line.

During the simulation the code is looking for shocks between triplets of
originally neighboring particles, which are given by the changes of the sign of
the volume spanned by these particles.  At shocks we attribute to particles
with velocity $\vec{v}$ a 'temperature', $ kT \approx \mu_M m_H (\vec{v} -
\vec{U})^2/3$, where $\vec U$ is the local velocity determined by interpolation
of the velocity field onto a twice coarser grid.  Subsequently, we suppose each
particle has properties both of dark and baryonic matter (we use a cosmic
abundance of $10 \%$ baryons).  The baryonic gas is allowed to cool and, during
gravitational collapse, it also gets adiabatically heated.  Further we take into
account a feedback mechanisms which results from star formation and subsequent
SN-explosions.  It reheats a certain part of the cold gas ($85 \%$, this
percentage was taken over from the older simulations).  This recipe leads to a
reasonable account of the condensation of baryonic matter in galactic halos and
its clustering.  Since we do not alter the dynamics during and after shocks,
dark and baryonic matter remains in the same pattern.

\midinsert
\vskip 1.5truecm
\par\noindent
\vglue7.2cm
\includegraphics{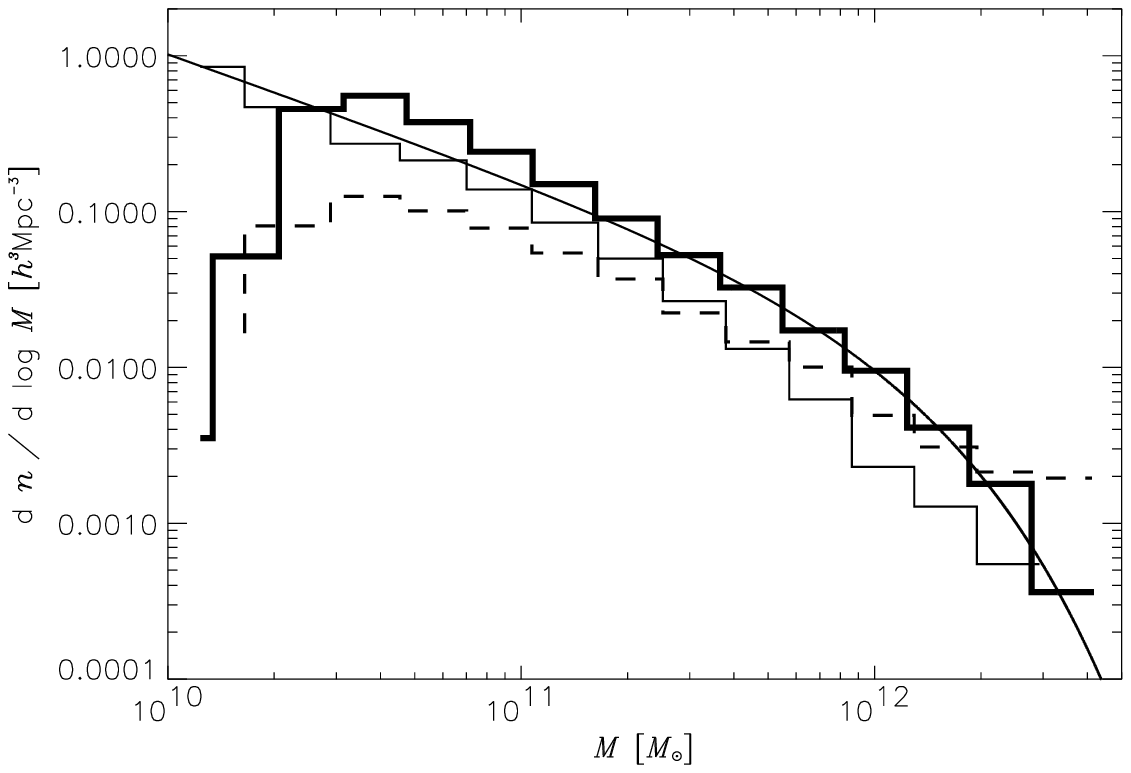}
\leftskip=1truecm
\rightskip=1truecm
\noindent
{\abstract FIGURE 2. The galaxy mass function of BSI-25 (solid histogram), 
BSI-25*(thick solid histogram), and CDM-25 (dashed histogram). The parameters 
of the fitting Schechter function are given in the text.}
\vskip -0.3truecm
\leftskip=0truecm
\rightskip=0truecm
\endinsert

Due to the nonlinear gravitational instability a large part of 'baryons' (up to
$88 \%$) gets 'cold' and transforms to star and star systems.  Galaxies are
identified with density maxima in the distribution of cold particles which lie
over a certain threshold (2 to 3 times the grid variance), but 'galaxy halos'
are assembled by all particles within a search radius of 0.5 times the mean
interparticle separation.  The resulting mass distributions can be fitted by a
Schechter function
$$
d n / d \log M = (n_* M / {M_*})^{-p} \exp (-M / M_* ) ,
$$
\ni For BSI-25* we get $M_*=10^{12}$ \Msun, $n_*=2.6 \times 10^{-2}
h^3\Mpc^{-3}$ and $p=0.8$.  This is a quite reasonable density of $M_*$ galaxies
(corresponding possibly to $L_*$ galaxies).  The derived galaxy catalogues are
used to compare the consequences of the model with characteristics of the large
scale matter distribution in the universe.

\midinsert
\vskip 1.5truecm
\par\noindent
\vglue7.2cm
\includegraphics{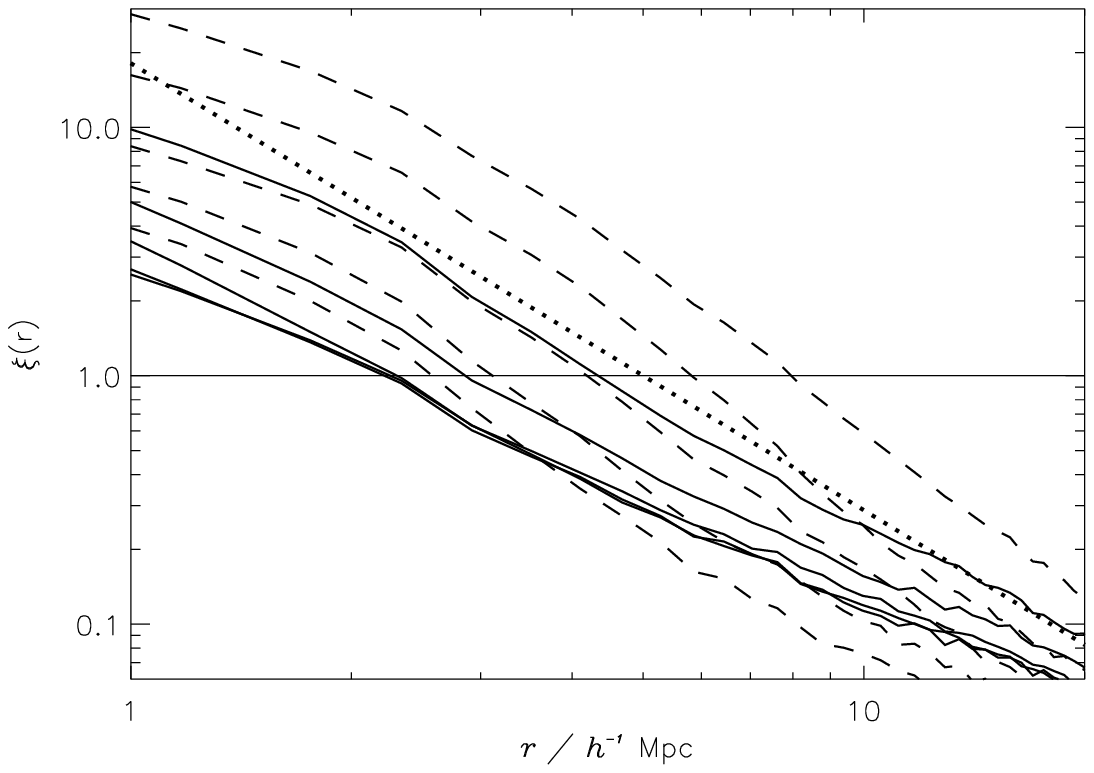}
\leftskip=1truecm
\rightskip=1truecm
\noindent
{\abstract FIGURE 3. Redshift evolution of the correlation function in BSI-75
and CDM-75, where on DM particle corresponds to $10^{11}$ \Msun.}
\vskip -0.3truecm
\leftskip=0truecm
\rightskip=0truecm
\endinsert

\smallskip
\ni III. POWER SPECTRA AND CORRELATION FUNCTION

The described galaxy identification scheme leads to a physical biasing model.
We used counts in cell variances as an robust measure of the large scale bias
(integral bias due to Cen and Ostriker, 1992) as a characteristic of the derived
galaxy catalogues.  We get in the BSI model 'naturally biased' galaxies with a
bias parameter $b \approx 1.5$ which does not strongly depend on the scales.  On
the other hand, 'clusters of galaxies', identified in the larger simulation
boxes, have $b \approx 3$ (Kates et al.  1995).  The question whether these
values of the bias parameter are realistic has been discussed using the redshift
dependence of the two-point correlation function.  Here, we used for
computational convenience particles in cells with overdensity of about 30 as
tracers of the galaxy distribution in BSI.  This threshold is an compromise
between the overdensity expected for virialised halos and the finite resolution
in the PM-simulations.  For the CDM simulations, we suppose no biasing at all.
In Fig.  3 we show the correlation functions at redshifts $z = 2, 1.5, 1, 0.5,
0$ (increasing from below) for BSI-75 (solid lines), and CDM-75 (dashed line).
The dotted line is about the observed curve
$$
\xi = ( r / r_0 )^{-1.8}
$$
with $r_0 = 5$ \hMpc.  As it is well known, the CDM curves are rising strongly
even in comoving coordinates during the redshift interval studied.  For the
chosen normalisation, the 'correct' amplitude and slope is reached about at
redshift $z \approx 1$.  In the contrary, for the BSI model we find between $z =
2$ and $z = 1$ a stable clustering in {\sl comoving coordinates}, and later an
increase of the slope and amplitude about up to the observed values.  In
Amendola et al.  (1995) we used these peak selected 'galaxy catalogues' for a
comparison with the APM angular correlation function, which is reasonably
reproduced in the BSI model.  Now we consider the comparison with galaxy
redshift catalogues using galaxies selected from density maxima in the cold gas
as described above (cp. Fig. 2).

\midinsert
\vskip 1.5truecm
\par\noindent
\vglue7.5cm
\includegraphics{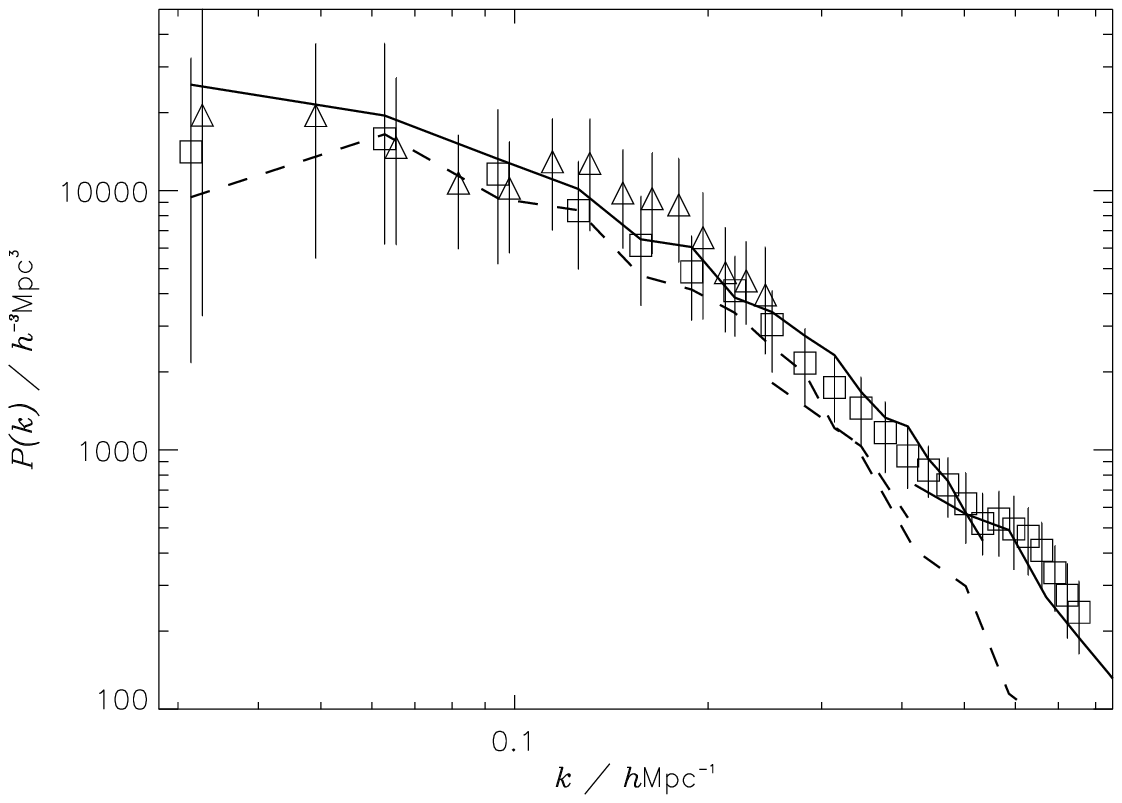}
\leftskip=1truecm
\rightskip=1truecm
\noindent
{\abstract FIGURE 4. Comparison with BSI (solid line) and CDM (dashed line)
galaxy power spectra from CfAII (squares) and IRAS (triangles).} 
\vskip -0.3truecm
\leftskip=0truecm
\rightskip=0truecm
\endinsert

First we describe the comparison of simulated power spectra with the power
spectrum reconstructed from the CfA galaxy catalogue (Vogeley \ea 1992).  We
impose redshift corrections by placing an arbitrary observer far outside the
simulation box.  The resulting power spectrum of BSI and CDM galaxies,
identified in simulations with a box length of 200 \hMpc is shown in Fig.  4,
where we select 'galaxies' with more than 10 particles for BSI and over 30
particles for CDM (the higher threshold is due to the more advanced clustering
in the CDM model).  The CDM model is strongly influenced by the redshift
corrections which enhance the power near the maximum, and they suppress the
power at higher $k$ values (another cause for suppression of the power at high
wave numbers is the subtraction of shot noise).  Both spectra can fit the CfA
power spectrum, in particular the BSI model leads to a very good fit of the
data.  The galaxies used in this simulation are biased with respect to the sea
of all dark matter particles by an (approximately linear) bias factor of about
2, contrary to the CDM model which has $b \approx 1$.

\midinsert
\vskip 1.5truecm
\par\noindent
\vglue7.3cm
\includegraphics{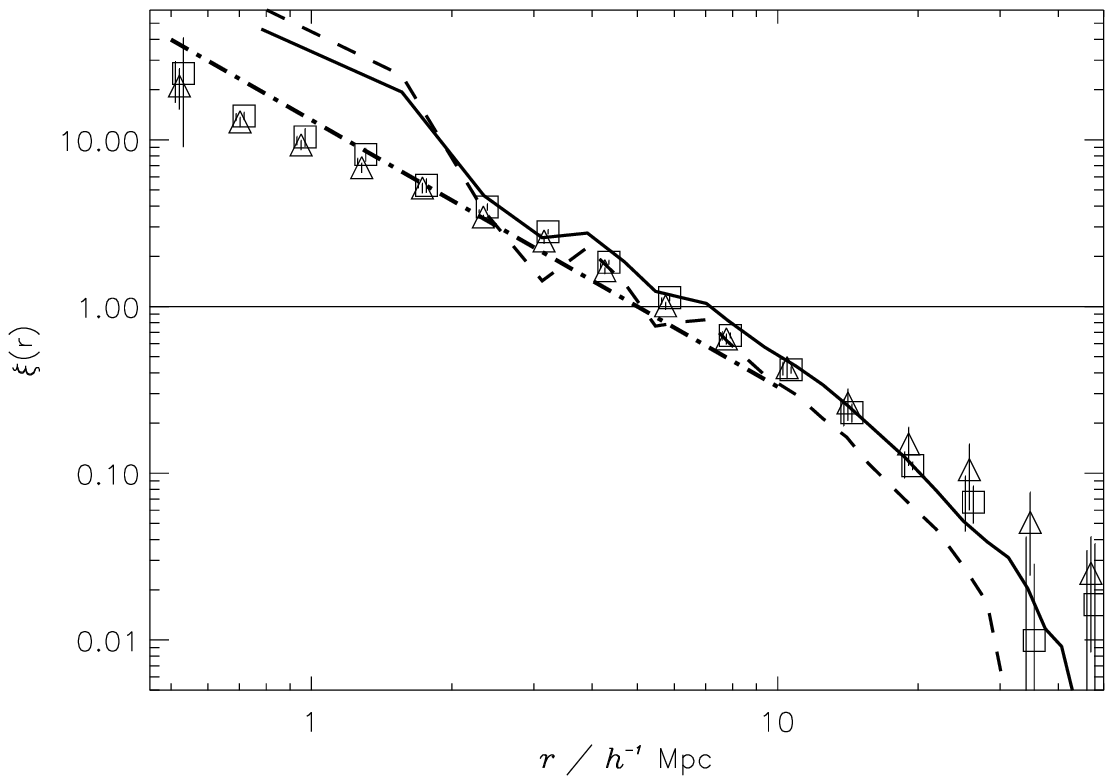}
\leftskip=1truecm
\rightskip=1truecm
\noindent
{\abstract FIGURE 5. Comparison with the two-point correlation in BSI (solid line) 
and CDM (dashed line) with results from the Las Campanas survey 
(triangles: number weighted, squares: luminosity weighted).}

\vskip -0.3truecm
\leftskip=0truecm
\rightskip=0truecm
\endinsert

The galaxy correlation function shown in Fig.  5 supplements the study of the
power spectrum.  While in principle it has the same information content, the
accuracy of its estimation is much higher in the region of nonlinear clustering.
Here we compare 'galaxies' composed of a minimum number of 5 and 10 particles in
the BSI-75 and CDM-75 simulations, respectively.  The data of the Las Campanas
Deep Redshift Survey are taken from Tucker \ea (1995).  Again, due to redshift
corrections both curves are able to reproduce the observed galaxy clustering
over a wide range of scales (in the contrary, the real space correlation
function of CDM galaxies is much steeper than that of BSI galaxies, cp.  Fig.
3).  The lower clustering level of the data at lengths below 2 \hMpc is probably
due to selection effects.  A small difference becomes visible in the length
range $(20 - 30)$ \hMpc, where the higher clustering of the LCDRS galaxies may
indicate an enhanced power, it is slightly better reproduced by the BSI model.
The dash-dotted line shows the power law $\xi = (r/r_0)^{-1.6}$ with $r_0 = 5$
\hMpc.

\smallskip
\ni IV. CONCLUSIONS

Strong differences between the properties of galaxies in the two models become
visible in the small scale velocity dispersion.  Its one dimensional (line of
sight) projection value lies at about 250 km/s in the BSI model and at about 600
km/s in the CDM model.  This velocity dispersion can be estimated from the
anisotropy of the two-point correlation function of galaxy redshift catalogues
with respect to the line of sight and orthogonal to it.  Further strong
differences come from the observed mass distribution of galaxy clusters and the
cluster-cluster correlation function (M\"uller 1994).  In the BSI model, the
latter remains positive on scales up to at least 60 \hMpc.  On the other hand,
the high degree of clustering in the CDM model is connected with large velocity
fields on small scales, which weakens the clustering signal if observed in
redshift space.  Therefore, the two-point correlation function and power spectra
have difficulties in distinguishing between the two models.  More elaborate
tests take into account the properties of galaxy clusters and large scale
velocity fields.  Complementary to the evolution of the correlation function,
both models can be clearly distinguished in the redshift evolution of the
characteristic scales of pancakes and filaments identified by more refined
statistical methods (Doroshkevich et al.  1995).

\smallskip
\vskip 0.1truecm
\ni {REFERENCES}
\ssk
\ref Gottl\"ober, S., M\"uller, V., Starobinsky, A.A., 1991, 
     {\sl Phys.Rev.}, D43, 2510.

\ref Kates, R., M\"uller, V., Gottl\"ober, S., M\"ucket, J.P.,
     Retzlaff, J., 1995, {\sl MNRAS} in press, astro-ph/9507036.

\ref Starobinsky, A.A., 1985, {\sl Sov.Astron.Lett.}, 9, 302.

\ref Semig, L., M\"uller, V., 1994, {\sl A}{\&}{\it A} in press, 
     astro-ph/9508020.

\ref Gottl\"ober, S., M\"ucket, 1993, {\sl A$\&{} $A} 272, 1.

\ref Kates, R., Kotok E., Klypin, A. 1991, {\sl A$\&{} $A} 243, 295.

\ref Cen, R., Ostriker, J.P., 1992, {\sl ApJ} {\bf 399}, L113.

\ref Amendola, L., Gottl\"ober, S., M\"ucket, J.P., M\"uller, V., 
     1995, {\sl ApJ} in press, astro-ph/9408104.

\ref Vogeley, M., Park, C., Geller, M., Huchra, J.P., 1992,  
     {\sl Astroph. J.} {\bf 391}, L5. 

\ref Tucker, D.L., Oemler, A.A., Kirshner, R.P., Lin, H.,
     Shectman, S.A., Landy, S.D., Schechter, P.L., 1995, 
     Proc. XXX Moriond-Meeting "Clustering in the Universe".}

\ref M\"uller, V., 1994, in: 'Cosmological Aspects of X-Ray Clusters
     of Galaxies', ed. W.C. Seitter, Kluver, NATO ASI Series, 441.

\ref Doroshkevich, A.G., Fong, R., Gottl\"ober, S., M\"ucket, J.P., 
     M\"uller, V., 1995, {\sl MNRAS} subm, astro-ph/9405088.

\end